# Navigator-Free Submillimeter Diffusion Imaging using Multishot-encoded Simultaneous Multi-slice (MUSIUM)

Wei-Tang Chang[1,2], Khoi Minh Huynh[1], Pew-Thian Yap[1,2], Weili Lin[1,2]

[1] Biomedical Research Imaging Center, University of North Carolina at Chapel Hill, NC, USA

[2] Department of Radiology, University of North Carolina at Chapel Hill, NC, USA

*Corresponding author at: **Biomedical Research Imaging Center,** University of North Carolina at Chapel Hill, NC, USA. Fax: +1 (919) 843-4456.
E-mail address: weitang_chang@med.unc.edu (W.-T. Chang).

Key words:    SMS, ultrahigh resolution, dMRI, multishot

## ABSTRACT

The ability to achieve submillimter isotropic resolution diffusion MR imaging (dMRI) is critically important to study fine-scale brain structures, particularly in the cortex. One of the major challenges in performing submillimeter dMRI is the inherently low signal-to-noise ratio (SNR). While approaches capable of mitigating the low SNR in high resolution dMRI have been proposed, namely simultaneous multi-slab (SMSlab) and generalized slice dithered enhanced resolution with simultaneous multislice (gSlider-SMS), limitations are associated with these approaches. The SMSlab sequences suffer from the slab boundary artifacts and require additional navigators for phase estimation. On the other hand, gSlider sequences require relatively high RF power and peak amplitude, which increase the SAR and complicate the RF excitation. In this work, we developed a navigator-free multishot-encoded simultaneous multi-slice (MUSIUM) imaging approach on a 3T MR scanner, achieving enhanced SNR, low RF power and peak amplitude, and being free from slab boundary artifacts. The dMRI with ultrahigh resolution (0.86 mm isotropic resolution), whole brain coverage and ~12.5 minute acquisition time were achieved, revealing detailed structures at cortical and white matter areas. The simulated and in vivo results also demonstrated that the MUSIUM imaging was minimally affected by the motion. Taken together, the MUSIUM imaging is a promising approach to achieve submillimeter diffusion imaging on 3T MR scanners within clinically feasible scan time.

## INTRODUCTION

Diffusion MRI (dMRI) is a noninvasive imaging modality that enables characterization of tissue microstructure as well as structural connectivity of the human brain (Basser et al., 1994). While massive diffusion datasets have been collected with a spatial resolution of 1.25-mm isotropic in the Human Connectome Project (Sotiropoulos et al., 2013), the ability to acquire ultra-high isotropic resolution diffusion images is critically important to study fine-scale brain structures and empowers the assessments of early brain development. However, the major impediments enabling submillimeter dMRI include long acquisition time and inherently low signal-to-noise ratio (SNR). Over the past decade, the simultaneous multislice (SMS) imaging with blipped-controlled aliasing (blipped-CAIPI SMS) has become one of the most widely-employed approaches to shorten acquisition time (Setsompop et al., 2012). While the SMS approach is capable of shortening acquisition time considerably, the low SNR remains an issue hindering our ability to achieve submillimeter dMRI.

Increasing both excitation volume per shot and number of shots are common strategies to enhance the SNR. Two different approaches have leveraged this concept, namely 3D multislab dMRI and RF-encoded multishot dMRI, respectively. The 3D multislab dMRI excites thick slabs sequentially in each TR, and Fourier encoding is performed along the slab direction for each slab over multiple TRs (Frank et al., 2010, Engstrom and Skare, 2013, Frost et al., 2015). Combining 3D multislab and SMS imaging, the simultaneous multi-slab (SMSlab) approaches enable rapid acquisition while preserving SNR (Chen and Feinberg, 2013, Frost et al., 2013, Bruce et al., 2017). Nevertheless, the aforementioned approach suffers from a few drawbacks. First, imperfect slab-selective profiles could lead to slab boundary artifacts. Second, the use of Fourier encoding makes it difficult to self-navigate the motion-induced phase variation across shots when the encoding frequency is high. Third, Gibbs artifacts could be present when a small slab encoding number is used. However, higher number of slab-encoding shots increases the slab thickness and therefore shortens the TR per shot, limiting its ability to employ a TR achieving the highest SNR efficiency (Chen and Feinberg, 2013). A number of innovative methods have been proposed to address these issues. For example, the nonlinear inversion for slab profile encoding was proposed to mitigate slab boundary artifacts but at the cost of a high computational load (Van et al., 2015, Wu et al., 2016). Besides, the 2D navigator was used to capture the phase variation (Bruce et al., 2017) of SMSlab. Nevertheless, the insertion of a 2D navigator increases the acquisition time and specific absorption rate (SAR).

In contrast to the 3D multislab approaches, the RF-encoded multishot approach, Generalized Slice Dithered Enhanced Resolution with Simultaneous Multislice (gSlider-SMS) (Setsompop et al., 2017), acquires multiple thin slabs simultaneously. Each thin slab consists of multiple thin slices that are RF-encoded, similar to the Hadamard encoding (Glover and Chang, 2012). Since RF-encoding is used, the gSlider approach offers two major advantages over the 3D mutlislab approaches, including a higher SNR to allow phase self-navigation and free from Gibbs ringing artifacts even if the number of within-slab slices is small. Nevertheless, the gSlider-SMS imposes several technical challenges including elevated SAR and RF peak amplitude. Although the RF peak amplitude can be reduced by variable-rate selective excitation (VERSE) (Hargreaves et al., 2004), it renders the gSlider imaging sensitive to B1 and B0 inhomogeneity. Secondly, in order to minimize slice crosstalk, it requires a long TR ($\geq$ 4s) (Setsompop et al., 2017), potentially compromising SNR efficiency.

To this end, we developed a multishot-encoded simultaneous multi-slice (MUSIUM) imaging approach, capable of achieving submillimeter dMRI on 3T without specialized gradient coils and mitigating the aforementioned limitations. Similar to the gSlider, MUSIUM excites a multiple of SMS factor slices and employs multiple shots to maintain an effective slice acceleration rate. However, differing from the gSlider, the simultaneously-excited slices in the MUSIUM sequences are equally spaced instead of forming multiple slabs, mitigating the need of a high BWTP in order to reduce SAR. To unalias the simultaneously-excited slices, the MUSIUM sequences use Fourier encoding (Zahneisen et al., 2013) rather than RF encoding so that the phase-optimization schemes (Wong, 2012) can be applied and thereby reduce the RF peak amplitude considerably. Unlike the SMSlab sequences in which most of the Fourier encoding shots are related to high spatial frequency, the encoding shots of MUSIUM imaging correspond to relatively low spatial frequency. As a result, each encoding shot achieves a high SNR, enabling phase self-navigation. Our results showed that the MUSIUM approach achieved a higher SNR when compared to conventional SMS imaging and was capable of acquiring whole-brain dMRI with 0.86-mm isotropic resolution and 66 diffusion directions in ~12.5 minutes. The motion robustness of MUSIUM imaging was also demonstrated.

**THEORY**

**MUltishot-encoded SImUltaneous Multi-slice (MUSIUM)**

The slice excitation and k-space sampling trajectory are illustrated in Figure 1a and 1b respectively. Let $n_{sms}$ denotes the effective SMS factor and $n_{shot}$ denotes the number of Fourier encoding shots. As shown in Figure 1a, $n_{sms} \times n_{shot}$ slices will be excited simultaneously for each shot. The fully sampled k-space of SMS can be viewed as the Fourier encoding of a 3D volume comprising of $n_{sms} \times n_{shot}$ contiguous slabs that are centered around the excited slices (dashed lines, Fig 1a) (Zahneisen et al., 2013). Hence, SMS is applicable to the existing 3D encoding. The fully-sampled k-space points of Figure 1a are denoted by the orange circles in Figure 1b. Differing from the conventional 3D encoding, the SMS 3D encoding is virtually free from Gibbs ringing effect (Czervionke et al., 1988) regardless of the number of partition encoding (PAE). This is because the Gibbs artifact is associated with the truncation of Fourier series that represent continuous signal in the image domain. For the SMS 3D encoding, however, the image signal along the slice direction is discrete instead of continuous. Therefore, the Fourier transform of the image signal can be approximated as discrete Fourier transform in which Gibbs phenomenon is not applicable.

To enhance SNR and reduce geometry factor (g-factor) penalty, the SMS excitation will be repeated by $n_{shot}$ times with different k-space sampling trajectories across shots as shown in Figure 1b. The solid dots of different colors denote the sampled $k_x$-lines of different shots. Every trajectory runs across not only the central but also the outer parts of k-space along the kz direction in order to facilitate the phase estimation of every slice. The sampling trajectories are designed to provide unique sampling points in k-space but also to have comparable image SNR with each other. The sampling trajectories of different shots start from different bands in the k-space. The boundaries of the bands are indicated by the dashed lines in Figure 1b. The width of the band along $k_z$ is related to the size of field-of-view (FOV) in the image domain. Let $FOV_z$ denotes the width of FOV along z direction. The width of the k-space band along $k_z$ is $1/d_{PAE}$, where $d_{PAE}$ is $FOV_z/n_{sms}$. The blip moment of y gradient ($G_y$) is $n_{ipat}/FOV_y$ where $n_{ipat}$ denotes the in-plane acceleration rate and $FOV_y$ denotes the width of FOV along y direction. Different from the blip moment of $G_y$, the blip moment of $G_z$ depends on the sampling location on the $k_y$-$k_z$ plane. Let $i_{trj}$ = 0, 1, 2 … denotes the index of the sampled $k_x$ lines in a trajectory, $k_z(i_{trj})$ denotes the $k_z$ position at the $i_{trj}$-th sample, $\Delta k_z(i_{trj})$ denotes the blip moment of $G_z$, $b_z(i_{trj})$ denotes the sign of the $\Delta k_z(i_{trj})$ and $p_z(i_{trj})$ denotes the index of k-space band. The shot index $i_{shot} = -\lfloor n_{shot}/2 \rfloor, 0,$ 1, … $(n_{shot}-\lfloor n_{shot}/2 \rfloor-1)$ where $\lfloor \cdot \rfloor$ denotes the floor function. The generation of every sampling trajectory is elaborated with pseudo programming language in Table 1.

1: Setup the initial $k_z(0) = (i_{shot} + 0.25)/d_{PAE}$ if $i_{shot}$ is even, and $(i_{shot} - 0.25)/d_{PAE}$ if $i_{shot}$ is odd.
2: Setup the initial $b_z(0) = -1$ if $i_{shot}$ is even, and $b_z(0) = 1$ if $i_{shot}$ is odd.
3: Setup the initial $p_z(0) = i_{shot} + \lfloor n_{shot}/2 \rfloor$
4. Set $i_{trj} = 0$
5. Repeat
6: if $p_z(i_{trj}) + b_z(i_{trj}) < 0$
   $\Delta k_z(i_{trj}) = -0.5 \times b_z(i_{trj})/d_{PAE}$, $p_z(i_{trj}+1)=0$ and $b_z(i_{trj}+1)=1$
  else if $p_z(i_{trj}) + b_z(i_{trj}) > (n_{shot}-1)$
   $\Delta k_z(i_{trj}) = -0.5 \times b_z(i_{trj})/d_{PAE}$, $p_z(i_{trj}+1) = n_{shot}-1$ and $b_z(i_{trj}+1) = -1$
  else
   $\Delta k_z(i_{trj}) = b_z(i_{trj})/d_{PAE}$, $p_z(i_{trj}+1) = p_z(i_{trj})+b_z(i_{trj})$ and $b_z(i_{trj}+1) = b_z(i_{trj})$
7: $k_z(i_{trj}+1) = k_z(i_{trj}) + \Delta k_z(i_{trj})$
8: set $i_{trj} = i_{trj}+1$
9: Until $i_{trj}$ reach the end of sampling trajectory

Table 1: The procedure for generating the blip moment of $G_z$ along the sampling trajectory.

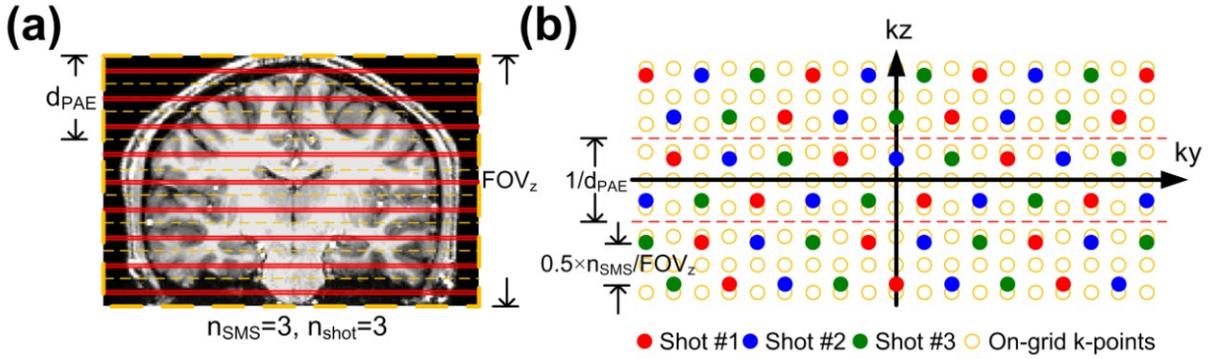

Figure 1: The RF excitation and k-space sampling trajectories of the MUSIUM sequences. (a) The RF excitation of a MUSIUM sequence. $n_{sms} \times n_{shot}$ slices are excited simultaneously. (b) The sampling trajectories of MUSIUM sequence.

**MATERIALS AND METHODS**

All experiments were performed using a 3T Prisma MR scanner (Siemens Healthcare, Erlangen, Germany). Data was acquired from three healthy male participants after obtaining informed consent using an institutionally approved protocol.

**Simultaneous Multislice Excitation**

The RF excitation and refocusing pulses were designed using the Shinnar-Le Roux (SLR) algorithm (Pauly et al., 1991). The limits of the ripple in the passband and stopband were set as 1%. To perform simultaneous multislice excitation, slice-selective RF pulses were frequency modulated and summed. In total, $n_{sms} \times n_{shot}$ slices were excited simultaneously. An optimized phase term for each of the summed RF pulses was introduced to reduce the peak RF amplitude by approximately square root of ($n_{sms} \times n_{shot}$) (Wong, 2012). The BWTP of excitation and refocusing pulses were 4 and 5.4, respectively. The pulse durations were 3.2 ms and 6.4 ms.

**Bloch Simulation of RF Load and Slice Crosstalk**

To compare between gSlider and our newly developed approach, the slice crosstalk, RF power and peak RF amplitude were obtained by simulating signal progression during spin-echo acquisition using Bloch simulation. The RF pulses of both gSlider and MUSIUM sequences were generated using SLR algorithm. T1 was set as 1400 ms (Lu et al., 2005, Stanisz et al., 2005, Wright et al., 2008) and TE as 80 ms. The T2 and T2* effects were ignored for simplicity. The pulse durations for excitation and refocusing were set as 6.4 and 8.0 ms, respectively. For the gSlider sequence, the BWTPs of excitation and refocusing pulses were 12 and 8, respectively as described in (Setsompop et al., 2017). For MUSIUM sequence, the BWTPs of excitation and refocusing pulses were 4.0 and 5.4, respectively.

The metrics of RF loading included the RF power and RF peak amplitude. The RF power was defined as the power integral of excitation and refocusing RF pulses divided by TR and the RF peak amplitude was the maximal magnitude of RF pulses across time. The slice cross talk measured the signal leakage from one of the simultaneously-acquired slices to the others, which was derived through the forward encoding matrix **A**. For gSlider imaging, the **A** contains the RF-encoding information obtained from the Bloch simulation. The individual slice information can be reconstructed by

$$\mathbf{z} = (\mathbf{A}^H\mathbf{A}+\lambda\mathbf{I})^{-1}\mathbf{A}^H\mathbf{b} = \mathbf{A}_{inv}\mathbf{b}, \tag{1}$$

where **z**, $\lambda$, and $\mathbf{A}_{inv}$ and **b** denote the vector of reconstructed slices, Tikhonov regularization parameter, the inverse matrix of **A** and the vector of encoded data across shots, respectively. The $\lambda$ was set as 0.4 as that in (Setsompop et al., 2017). The point spread function of the reconstruction can be characterized by impulse response analysis

$$\mathbf{r}_i = \mathbf{A}_{inv}\mathbf{A}\boldsymbol{\delta}_i \qquad (2)$$

where $\mathbf{r}_i$ denotes the vector of impulse response corresponding to slice i and $\boldsymbol{\delta}_i$ is the impulse vector. On the other hand, for MUSIUM imaging, the $\mathbf{A}$ is the approximately the Fourier encoding matrix and the $\mathbf{A}_{inv}$ in (2) is the inverse Fourier matrix. Let $r_{ij}$ denotes the elements in $\mathbf{r}_i$. The slice cross talk at slice i can be calculated as

$$\frac{\sqrt{\sum_{j \neq i}|r_{ij}|^2/(n_{shot}-1)}}{|r_{ii}|} . \qquad (3)$$

**Data Acquisition and Reconstruction**

Imaging studies were conducted to 1) compare the SNR between conventional SMS and MUSIUM sequences, and 2) to demonstrate the efficacy of ultrahigh resolution whole-brain dMRI. The SMS diffusion images were acquired using the following imaging parameters: matrix dimension=192×180×126 (R-L×H-F×A-P) mm$^3$; 1.0 mm isotropic resolution; axial slicing; SMS factor=2; excitation and refocusing RF pulse durations were 3.2 and 6.4 ms, respectively; BWTP was 4.0 and 5.4 for excitation and refocusing pulses, respectively; no in-plane acceleration; partial Fourier 5/8; echo spacing=0.93 ms; TE/TR=83 ms/11.43 s; b=2000 s/mm$^2$. The imaging parameters for MUSIUM were similar as those of SMS except for the following modifications: number of shots = 3; TR per shot = 3.81 s and effective TR per volume = 11.43 s.

To demonstrate the efficacy of obtaining ultrahigh resolution dMRI using MUSIUM, the imaging protocols were as follows: matrix dimension=256×208×153 (R-L×H-F×A-P) mm$^3$; 0.86 mm isotropic resolution; axial slicing; SMS factor = 3; number of shots = 3, excitation and refocusing RF pulses durations were 3.2 and 6.4 ms, respectively; BWTP was 4.0 and 5.4 for excitation and refocusing pulses, respectively; no in-plane acceleration; partial Fourier 5/8; echo spacing=0.97 ms; TE = 99 ms; TR per shot = 3.65 s and effective TR = 10.95 s. There were 5, 15, 45 diffusion directions at b=500, 1000, 2000 s/mm$^2$ respectively with 1 $b_0$ image. Total scan time was ~12.5 minutes.

The data acquired by MUSIUM sequences were reconstructed in two steps, namely phase estimation and SENSE-based reconstruction. Assuming a slowly varying in-plane phase, the phase information can be accurately approximated using low k-space information. Therefore, the images were down-sampled, which in turn provided a better SNR and much shorter reconstruction time. To estimate the shot-to-shot phase variation, the image of each shot was reconstructed. Let $y^{ds}$, $\mathbf{E}^{ds}$, and $x^{ds}$ denote the down-sampled data, down-sampled forward matrix and down-sampled image to be reconstructed, respectively.

$$y^{ds} = [y_1^{ds}, y_2^{ds}, \ldots y_{n_{shot}}^{ds}]^T, \quad (4)$$

$$x^{ds} = [x_1^{ds}, x_2^{ds}, \ldots x_{n_{shot}}^{ds}]^T, \quad (5)$$

$$\mathbf{E}^{ds} = \text{diag}(\mathbf{E}_1^{ds}, \mathbf{E}_2^{ds}, \ldots \mathbf{E}_{n_{shot}}^{ds}), \quad (6)$$

where $[\cdot]^T$ denotes the transpose operation. Let $i' = i_{shot} + \lfloor n_{shot}/2 \rfloor + 1$ denotes the shot index. The $\mathbf{E}_{i'}^{ds} = \mathbf{G}_{i'}^{ds} \cdot S^{ds}$ where $S^{ds}$ and $\mathbf{G}_{i'}^{ds}$ are the down-sampled coil sensitivity profile and down-sampled encoding matrix respectively for the *i'*-th shot, including k-space sampling trajectory, in-plane subsampling (if applicable), and slice summation. The coil sensitivity profile with original resolution was estimated from the central k-space lines of a reference image using ESPIRiT (Uecker et al., 2014). The total number of receiving coil is $n_{ch}$. The 2D kernel size was 7x7. The $x_i^{ds}$ encompasses the spatial information of $n_{sms} \times n_{shot}$ slices but $y_i^{ds}$ contains only one collapsed slice from $n_{ch}$ receiving coils. The reconstruction of $x^{ds}$ can be formulated as follows:

$$\widehat{x^{ds}} = \arg\min_{x^{ds}} \|y^{ds} - E^{ds} \cdot x^{ds}\|_2^2 + \lambda_{reg} \|D \cdot x^{ds}\|_2^2, \quad (7)$$

where D denotes the matrix of spatial differencing operation that promotes the in-plane smoothness. The first and second $\|\cdot\|_2^2$ are the data consistency and the regularization terms respectively. $\lambda_{reg}$ denotes the regularization parameter.

$$\lambda_{reg} = 0.1 \cdot \|y^{ds}\|_2^2 / \|D \cdot x_{ini}^{ds}\|_2^2, \quad (8)$$

where $x_{ini}^{ds}$ is the initial solution in (8) without the regularization term. In this study, only the central 1/4 $k_x$ and 1/4 $k_y$ points, i.e. 1/16 k-space, were used for phase estimation.

Although the images $\widehat{x^{ds}}$ are susceptible to undesirable noise amplification due to a high acceleration rate per shot, the shot-to-shot phase inconsistencies, which are expected to be spatially smooth, can be reliably estimated with the denoising operation based on total variation (Rudin et al., 1992).

$$e^{-j\widehat{\Phi}_i^{ds}} = TV(\hat{x}_i^{ds}) / |TV(\hat{x}_i^{ds})|, \quad (9)$$

where TV(·) denotes the total variation operation. The estimated phase images $e^{-j\hat{\Phi}_i^{ds}}$ were upsampled by interpolation using the matlab command 'interpn' with the 'spline' method. With the upsampled phase image $e^{-j\hat{\Phi}}$, the image to be reconstructed can be expressed as

$$x = e^{-j\hat{\Phi}} \cdot x_{mag} = \begin{bmatrix} e^{-j\hat{\Phi}_1} \\ e^{-j\hat{\Phi}_2} \\ \vdots \\ e^{-j\hat{\Phi}_{n_{shot}}} \end{bmatrix} \cdot x_{mag}, \qquad (10)$$

where $x_{mag}$ and $e^{-j\hat{\Phi}_{i'}}$ represent the magnitude image that is consistent across shots and the estimated phase image of the i'-th shot respectively. The reconstruction of $x_{mag}$ is similar to but slightly different from eq. (7):

$$\hat{x}_{mag} = \arg\min_{x_{mag}} \|y - E \cdot x_{mag}\|_2^2 + \lambda_{mag} \|x_{mag}\|_2^2, \qquad (11)$$

where y denotes the measured data with all the shots and $E=[E_1, E_2, ... E_{n_{shot}}]^T$ denotes the forward matrix. The $E_{i'} = G_{i'} \cdot e^{j\hat{\Phi}_{i'}} \cdot S$ where S denotes the coil sensitivity profile and $E_{i'}$ denotes the encoding matrix of the i'-th shot. The regularization parameter $\lambda_{mag}$ can be calculated as

$$\lambda_{mag} = \|y\|_2^2 / \|x_{ini}\|_2^2, \qquad (12)$$

where $x_{ini}$ is the initial solution in eq. (11) without the regularization term. The eq. (11) was solved by the conjugate gradient method using the MATLAB function 'pcg'. It can be seen from eq. (11) that the number of unknowns along z direction is $n_{sms} \times n_{shot}$ and the number of equations is $n_{shot}$. Since the effective slice acceleration remains at $n_{sms}$, the g-factor penalty of MUSIUM is comparable to that of the conventional SMS approach. Yet, the MUSIUM imaging has the SNR advantage over the SMS approach owing to the use of multi-shot acquisition.

**Data Denoising**

To further boost the SNR, a denoising method (MPPCA) based on the principal component analysis (PCA) and Marchenko-Pastur distribution was used (Veraart et al., 2016). The MPPCA method applies PCA denoising to overlapping spatial patches and chooses the rank value automatically based on the Marchenko-Pastur distribution. The patch size used in this study was 5x5x5. The MPPCA method was implemented using an in-house MATLAB code.

**Motion**

The proposed MUSIUM sequences capable of acquiring a 1-mm or submillimeter diffusion volume in ~10 s. Nevertheless, significant head motion may still occur within 10 sec in some particular groups, such as pediatric population, Parkinson's Disease patients, and so forth. Hence, it is important to assess the effects of such motion of acquired DTI images. We used a reference image with 1-mm isotropic resolution as the ground truth and tested the following two motion cases: i) linear translation of 1 mm in y, z axes and linear rotation of 1° in x, and ii) linear translation of 2 mm in y, z axes and linear rotation of 2° in x during the MUSIUM encoding acquisition. The x, y, z axes represent the readout, phase encoding, and slice directions. For axial slicing, the motion pattern is up-center-down as in "yes". The motion effects were simulated and the reconstruction was conducted as the proposed reconstruction framework without the knowledge of motion. The motion effects were also assessed in vivo. The two motion cases as used in the simulation were tested. The subject was prompted to move the head continuously center-up-down-center as in "yes" at 12-s interval during the scan. To facilitate the subject to produce small and large head rotations, three equally-spaced crosshairs were projected at the center, below the center and above the center of the screen. The participant was instructed to stare at the crosshair below the center during the "stay still" run. In the "small motion" run, the subject moved the head by changing the eye fixation between the crosshairs at and below the center of the screen. Likewise, the eye fixation was changed between the crosshairs below and above the center of the screen in the "large motion" run.

**RESULTS**

**Comparisons of RF Load and Slice Crosstalk by Bloch Simulation**

Using the same SMS factor and number of encoding shots, the RF-encoded approach as used in gSlider sequences and the Fourier-encoded approach as used in MUSIUM sequences excited an identical number of slices but in different ways (Figure 2a). For a fair comparison, none of the sequences used VERSE in the Bloch simulation. Figure 2b shows the comparisons of RF power, peak RF amplitude, $G_z$ strength and slice cross-talk between RF-encoded and Fourier-encoded approaches, respectively. The RF power ($\propto$SAR) and peak amplitude of RF-encoded approach is 13.4 and 10.8 times higher than that of Fourier-encoded approach, respectively. The maximally achievable RF amplitude depends on the specific RF hardware. If the RF amplitude exceeds the hardware limit, the delivered RF pulse could be clipped and the excitation profile could be erroneous. The required slice gradient strength of RF-encoded approach is 2.9 times higher than that of Fourier-encoded approach. The slice crosstalk of RF-encoded approach increases as the TR decreases while that of Fourier-encoded approach remains constantly small regardless of the TR. The smallest slice crosstalk of RF-encoded approach is 12.3 times higher than that of Fourier-encoded approach.

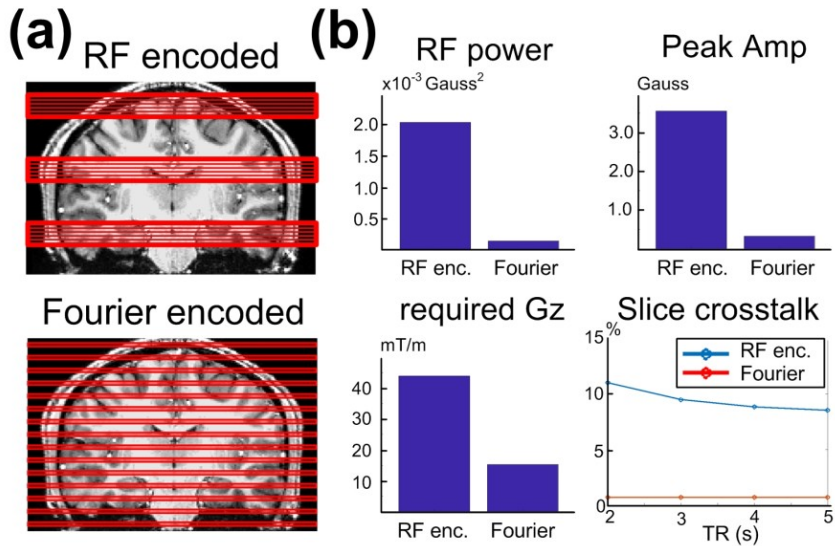

Figure 2: The comparisons between RF-encoded and Fourier-encoded approaches. (a) The illustrations of the slice excitations in RF-encoded and Fourier-encoded approaches, respectively. The effective SMS factor = 3, number of encoding shots = 5. (b) The comparisons of RF power, RF peak amplitude, required $G_z$ and the slice crosstalk. Abbreviations: Amp – Amplitude; enc. – encoding.

**SNR Enhancement by Multi-shot Diffusion Imaging**

The SNR comparison between single-shot SMS and multi-shot MUSIUM acquisitions is demonstrated in Figure 3. The number of shots in MUSIUM sequences is 3. Figure 3a showed the SNR maps of SMS and MUSIUM and the SNR ratio of MUSIUM to SMS is shown in Figure 3b. In general, the SNR is enhanced by the MUSIUM sequences when compared to the SMS. The averaged SNR of MUSIUM imaging is 56.2% higher than that of SMS imaging.

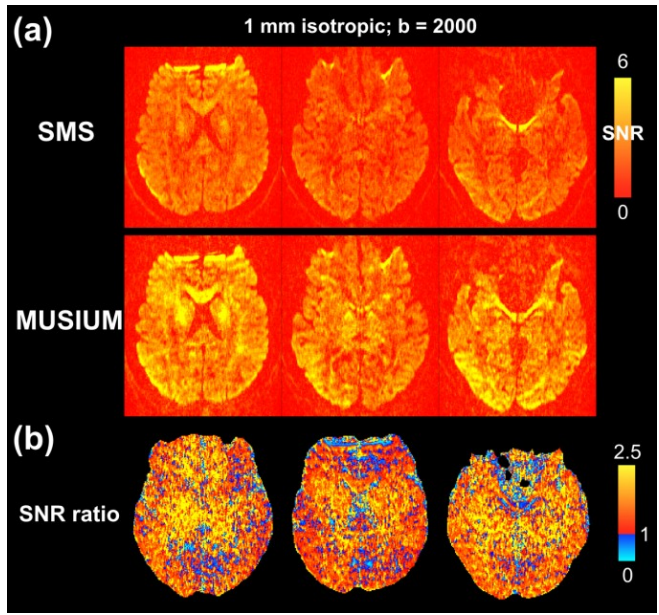

Figure 3: The SNR comparison between single-shot SMS and multi-shot MUSIUM acquisitions. (a) The SNR maps of SMS and MUSIUM acquisitions. (b) The SNR ratio of MUSIUM to SMS. The b value of the diffusion images is 2000. The spatial resolution is 1-mm isotropic. The SNR values and ratios are color-encoded as indicated by the color bars.

**Efficacy of sub-millimeter MUSIUM Imaging**

The efficacy of submillimeter MUSIUM imaging is demonstrated in Figure 4. Representative diffusion images of different b values are shown in Figure 4a. The color-FA results are shown in Figure 4b with the zoom-in tensor plot on the bottom-right. Although the tensors on the cortex as indicated by the opaque red arrow are a bit spatially noisy, they are generally perpendicular to the cortical surface.

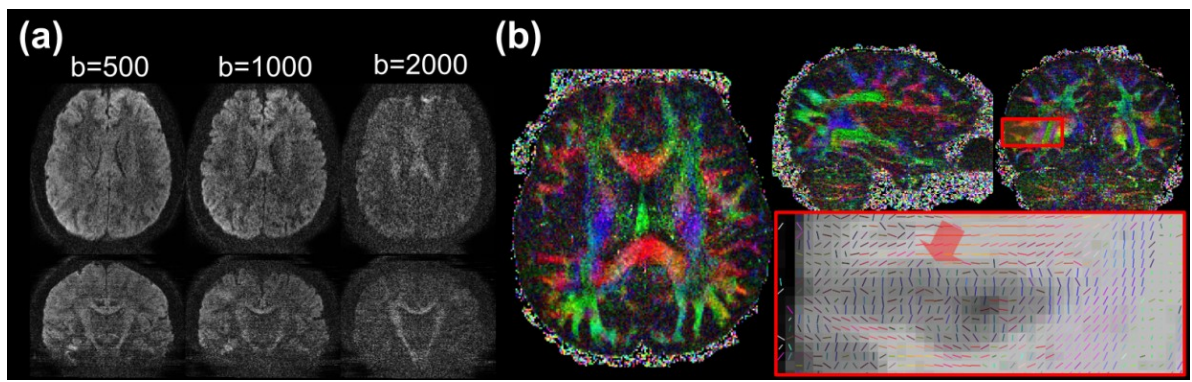

Figure 4: The reconstruction results of the 0.86-mm isotropic MUSIUM data. (a) The diffusion images at b=500, 1000 and 2000. (b) The FA map in three orthogonal planes. The data acquisition time is ~12.5 minutes.

Using the MPPCA denoising method, the signal quality can be further improved. Figure 5 demonstrated the denoised color-FA maps of three different subjects. The three zoom-in boxes highlighted by the red rectangles are centered around a similar location and the enlarged tensor plots are shown at the bottom row. In general, the tensor orientations are spatially stable and consistent across subjects.

The results demonstrate the ability of MUSIUM imaging to detect the fine-scale structures with the data acquisition time of ~12.5 minutes.

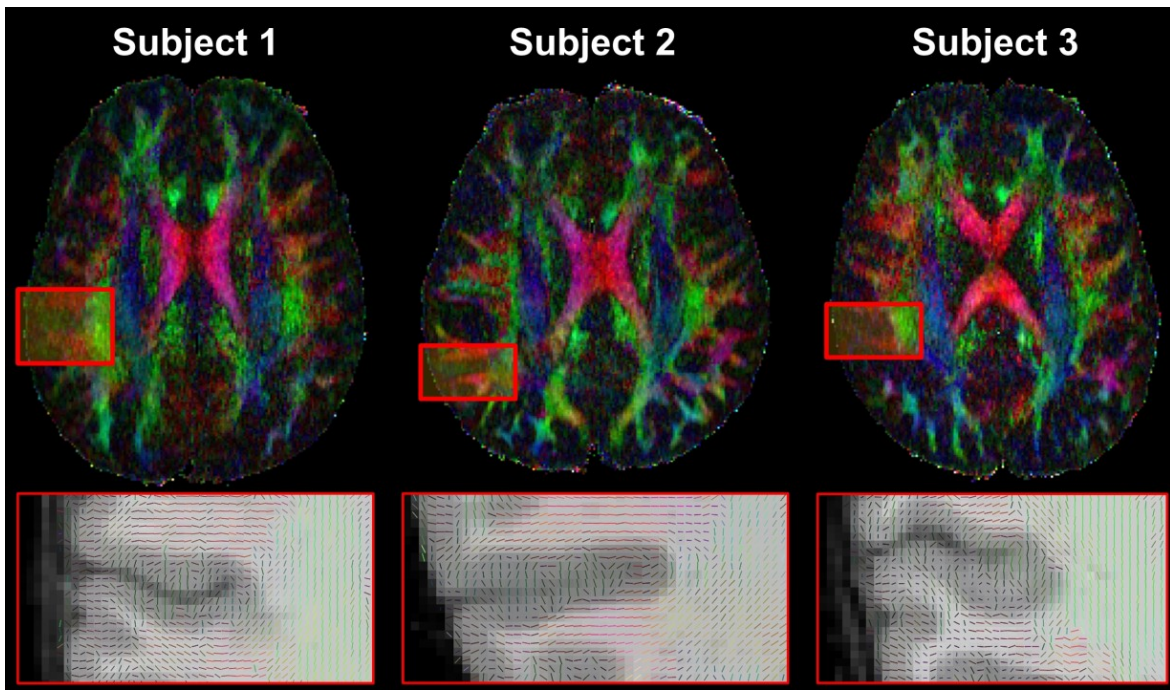

Figure 5: The denoised colored FA maps and zoom-in tensor plots of 0.86 mm isotropic data acquired from three different subjects. The tensor plots in the regions highlighted by the red box are enlarged in the bottom of the panel.

**Motion**

The effects of the motion along the MUSIUM encoding shots are shown in Figure 6. The upper row shows three different motion conditions. The middle and bottom rows show the simulated and in vivo results, respectively. The reconstructed images without motion are shown in Figure 6a serving as the reference. The linear translation of 1 mm and linear rotation of 1° during MUSIUM encoding lead to slight stripping artifacts in the reconstructed images as shown in Figure 6b. The artifacts in Figure 6c become more pronounced when the translation and rotation increase by two-fold. The stripping artifacts are also observed in the in vivo data and become more pronounced as the motion increased. Nevertheless, the general features of images are preserved despite the presence of such motions.

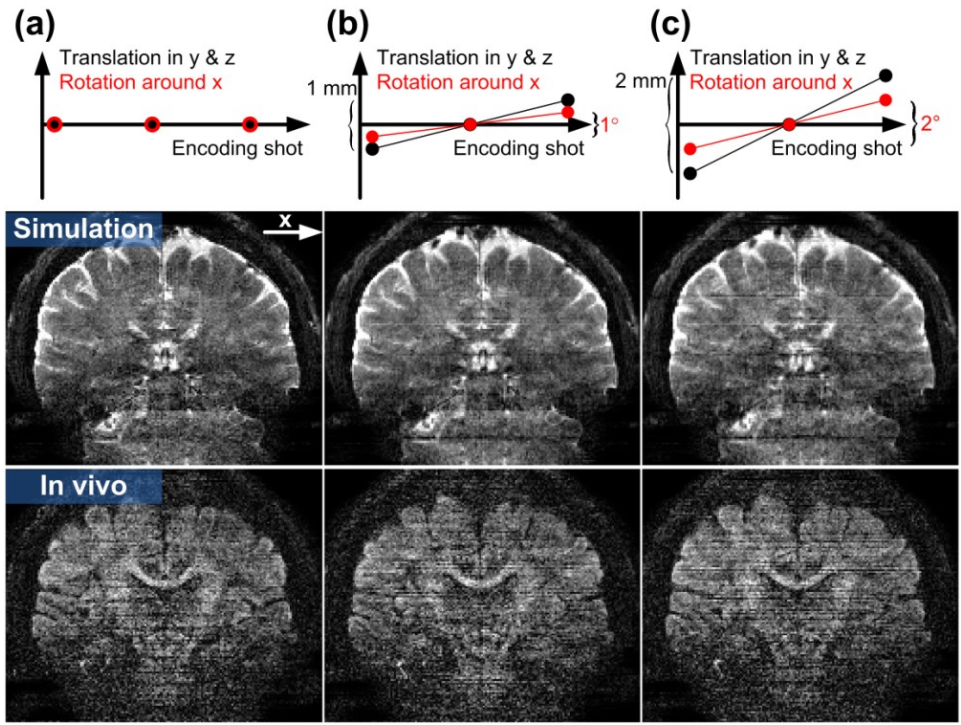

Figure 6: The motion effects on the MUSIUM reconstruction. Three motion conditions along the 3 encoding shots are simulated: (a) No motion, (b) linear translation of 1 mm in y and z axes and linear rotation of 1° around x axis, and (c) linear translation of 2 mm in y and z axes and linear rotation of 2° around x axis. The simulated and in vivo images are shown in the middle and bottom rows of each panel respectively.

**DISCUSSION**

A novel approach capable of acquiring ultra-high isotropic resolution, whole brain DTI images within a clinically acceptable acquisition time was proposed. The proposed MUSIUM imaging approach demonstrated several major advantages over existing approaches. A large number of slices were excited simultaneously in the MUSIUM sequences and multi-shot acquisitions were employed to enhance the SNR and reduce the g-factor penalty. The k-space sampling trajectories were designed in a way that the SNR of each shot is sufficient for phase estimation. Comparing to gSlider with an identical number of simultaneously-excited thin slices, the MUSIUM imaging showed a lower RF load, gradient strength and slice crosstalk. That is, MUSIUM imaging can be implemented on MR scanners without increasing the RF pulse duration, dedicated high performance gradient coils and the use of complicated RF excitation schemes such as VERSE. Compared to the single-shot SMS acquisition, in vivo MUSIUM images showed higher SNR efficiency. The efficacy of submillimeter MUSIUM imaging was demonstrated successfully at 0.86 mm isotropic resolution, revealing detailed structures at cortical and white matter areas.

**Acquisitions and reconstructions**

The k-space sampling trajectories of MUSIUM travel between $-k_z$ and $k_z$ with generally a larger blip size than conventional SMS, which may affect the echo spacing and TE. In the case of $n_{sms}=3$ and $n_{shot}=3$, the maximal blip moment in MUSIUM sequences doubles the blip moment of the blip-CAIPI SMS sequences with FOV/2 shift. Nevertheless, such a blip moment leads to a negligible increase of TE. Moreover, even if a shorter echo spacing can be achieved in SMS imaging, it might not be applicable to human imaging due to potential peripheral nerve simulation. Therefore, the influences of the MUSIUM sampling trajectories on the TE and echo spacing are minimal.

The selection of reconstruction approach is commonly a trade-off between signal quality, robustness and computational load. This study employed the SENSE-based method and the two-step approach based on the linear inverse estimation and relatively-low computational load. The MATLAB code ran on shared servers across the campus. With the requested 12 physical cores and 50-GB RAM, the reconstruction per diffusion volume takes ~18 minutes at the isotropic resolution of 0.86 mm. The computation time can be further shortened with parallel processing such as Techila Distributed Computing (Techila Technologies Ltd, Tampere, Finland).

Other reconstruction approaches such as low-rank matrix completion (Haldar, 2014, Mani et al., 2017, Mani et al., 2020), SNR-enhancing joint reconstruction (Haldar et al., 2013), nonlinear phase estimation (Haldar et al., 2020, Ramos-Llorden et al., 2020) have been proposed to improve signal quality at the costs of a high computational load. Nevertheless, the low-rank approaches reported the successful reconstruction on either in-plane acceleration data or through-plane acceleration data with low SMS factor (Mani et al., 2017, Mani et al., 2020). We also examined the reconstruction quality using the low-rank reconstruction approach on through-plane acceleration data with a high SMS factor (Figure S1). The results shown in Figure S2 suggested that the low-rank reconstructions of MUSIUM data are prone to introduce stripping artifacts in coronal/sagittal views for an axial acquisition and sensitive to the choices of regularization parameters. Nevertheless, other low-rank-based reconstruction, such as locally low-rank regularization (Hu et al., 2019), might improve the reconstruction quality over the low-rank approaches evaluated. However, the optimization of the reconstruction method is beyond the scope of our study.

The denoising techniques and multi-shot MR sequences are complementary approaches for SNR enhancement. The denoising techniques may fail if the SNR is too low as in high-resolution diffusion MRI because the covariance matrix of the signal needs to have a sufficiently low rank. Therefore, the enhanced SNR per shot provided by MUSIUM sequences is beneficial to the subsequent denoising process. In this study, we used the MPPCA approach to boost up the SNR on the diffusion analysis. While the MPPCA method effectively suppresses the noise, many recently proposed denoising methods could further improve the signal quality. An extension of the MPPCA method was proposed to apply on the noise models other than additive white Gaussian noise (Cordero-Grande et al., 2019). In addition, joint reconstructions of multiple diffusion-encoding directions were proposed to improve the SNR significantly despite the requirement of lengthy reconstruction time and much more memory (Haldar et al., 2020, Ramos-Llorden et al., 2020). Future work will explore the optimal trade-off between the SNR enhancement and computational load.

**Motion**

The motion effects during the MUSIUM acquisitions were tested by simulation and in vivo studies. Both simulated and in vivo results showed the motion introduced stripping artifacts, but the general image features were preserved. The robustness of MUSIUM sequences against motion can be expected because the sampling trajectories of each shot travel through the central part of k-space. Future work will perform the retrospective motion correction by using the reconstructed single-shot images for motion estimation and incorporating the motion information into the multi-shot reconstruction. Meanwhile, the markerless prospective motion correction will also be explored (Frost et al., 2019, Berglund et al., 2020).

**Limitations**

The accuracy of phase estimation is critical for the two-step reconstruction used in this study. The challenges for the phase estimation of MUSIUM imaging are the high acceleration rate per shot. The combined acceleration rate ($n_{sms} \times n_{shot}$) per shot in MUSIUM imaging is on the order of 10 if the in-plane acceleration is not employed, which is around the limit of parallel-imaging reconstruction capability. However, the combined acceleration rate could easily exceed such a limit if the in-plane acceleration is used. This imposes a major obstacle for MUSIUM imaging to achieve higher resolution. Ongoing efforts in our group have been carried out to tackle this issue.

**CONCLUSION**

The navigator-free MUSIUM sequences demonstrate enhanced SNR efficiency of dMRI compared to conventional SMS imaging. With the same SMS factor and number of shots, MUSIUM sequences demonstrate lower RF power, RF peak amplitude and slice crosstalk than the RF-encoded multishot approach. The in vivo MUSIUM images demonstrate the efficacy of submillimeter dMRI, revealing fine-scale structures at the cortex. In addition, the sophisticated design of k-space sampling trajectory renders the MUSIUM imaging less sensitive to motion. Taken together, the proposed MUSIUM imaging is a promising approach to achieve submillimeter diffusion imaging on general-purpose 3T MR scanners within clinically feasible scan time.

**ACKNOWLEGEMENTS**

We thank Kim-Han Tung and Ye Wu for providing helpful comments. This work was supported in part by NIH grants, R21AG060324 and U01MH110274.

**FIGURE DESCRIPTIONS**

Figure 1: The RF excitation and k-space sampling trajectories of the MUSIUM sequences. (a) The RF excitation of a MUSIUM sequence. $n_{sms} \times n_{shot}$ slices are excited simultaneously. (b) The sampling trajectories of MUSIUM sequence.

Figure 2: The comparisons between RF-encoded and Fourier-encoded approaches. (a) The illustrations of the slice excitations in RF-encoded and Fourier-encoded approaches, respectively. The effective SMS factor = 3, number of encoding shots = 5. (b) The comparisons of RF power, RF peak amplitude, required $G_z$ and the slice crosstalk. Abbreviations: Amp – Amplitude; enc. – encoding.

Figure 3: The SNR comparison between single-shot SMS and multi-shot MUSIUM acquisitions. (a) The SNR maps of SMS and MUSIUM acquisitions. (b) The SNR ratio of MUSIUM to SMS. The b value of the diffusion images is 2000. The spatial resolution is 1-mm isotropic. The SNR values and ratios are color-encoded as indicated by the color bars.

Figure 4: The reconstruction results of the 0.86-mm isotropic MUSIUM data. (a) The diffusion images at b=500, 1000 and 2000. (b) The FA map in three orthogonal planes. The data acquisition time is ~12.5 minutes.

Figure 5: The denoised colored FA maps and zoom-in tensor plots of 0.86 mm isotropic data acquired from three different subjects. The tensor plots in the regions highlighted by the red box are enlarged in the bottom of the panel.

Figure 6: The motion effects on the MUSIUM reconstruction. Three motion conditions along the 3 encoding shots are simulated: (a) No motion, (b) linear translation of 1 mm in y and z axes and linear rotation of 1° around x axis, and (c) linear translation of 2 mm in y and z axes and linear rotation of 2° around x axis. The simulated and in vivo images are shown in the middle and bottom rows of each panel respectively.

images without phase estimation through locally low-rank regularization. Magn Reson Med 81:1181-1190.

Lu H, Nagae-Poetscher LM, Golay X, Lin D, Pomper M, van Zijl PC (2005) Routine clinical brain MRI sequences for use at 3.0 Tesla. J Magn Reson Imaging 22:13-22.

Mani M, Jacob M, Kelley D, Magnotta V (2017) Multi-shot sensitivity-encoded diffusion data recovery using structured low-rank matrix completion (MUSSELS). Magn Reson Med 78:494-507.

Mani M, Jacob M, McKinnon G, Yang B, Rutt B, Kerr A, Magnotta V (2020) SMS MUSSELS: A navigator-free reconstruction for simultaneous multi-slice-accelerated multi-shot diffusion weighted imaging. Magn Reson Med 83:154-169.

Pauly J, Le Roux P, Nishimura D, Macovski A (1991) Parameter relations for the Shinnar-Le Roux selective excitation pulse design algorithm [NMR imaging]. IEEE Trans Med Imaging 10:53-65.

Ramos-Llorden G, Ning L, Liao C, Mukhometzianov R, Michailovich O, Setsompop K, Rathi Y (2020) High-fidelity, accelerated whole-brain submillimeter in vivo diffusion MRI using gSlider-spherical ridgelets (gSlider-SR). Magn Reson Med 84:1781-1795.

Rudin L, Osher S, Fatemi E (1992) Nonlinear total variation based noise removal algorithms. Physica D Nonlinear Phenomena 60:259-268.

Setsompop K, Fan Q, Stockmann J, Bilgic B, Huang S, Cauley SF, Nummenmaa A, Wang F, Rathi Y, Witzel T, Wald LL (2017) High-resolution in vivo diffusion imaging of the human brain with generalized slice dithered enhanced resolution: Simultaneous multislice (gSlider-SMS). Magn Reson Med.

Setsompop K, Gagoski BA, Polimeni JR, Witzel T, Wedeen VJ, Wald LL (2012) Blipped-controlled aliasing in parallel imaging for simultaneous multislice echo planar imaging with reduced g-factor penalty. Magn Reson Med 67:1210-1224.

Sotiropoulos SN, Jbabdi S, Xu J, Andersson JL, Moeller S, Auerbach EJ, Glasser MF, Hernandez M, Sapiro G, Jenkinson M, Feinberg DA, Yacoub E, Lenglet C, Van Essen DC, Ugurbil K, Behrens TE, Consortium WU-MH (2013) Advances in diffusion MRI acquisition and processing in the Human Connectome Project. Neuroimage 80:125-143.

Stanisz GJ, Odrobina EE, Pun J, Escaravage M, Graham SJ, Bronskill MJ, Henkelman RM (2005) T1, T2 relaxation and magnetization transfer in tissue at 3T. Magn Reson Med 54:507-512.

Uecker M, Lai P, Murphy MJ, Virtue P, Elad M, Pauly JM, Vasanawala SS, Lustig M (2014) ESPIRiT--an eigenvalue approach to autocalibrating parallel MRI: where SENSE meets GRAPPA. Magn Reson Med 71:990-1001.

Van AT, Aksoy M, Holdsworth SJ, Kopeinigg D, Vos SB, Bammer R (2015) Slab profile encoding (PEN) for minimizing slab boundary artifact in three-dimensional diffusion-weighted multislab acquisition. Magn Reson Med 73:605-613.

Veraart J, Novikov DS, Christiaens D, Ades-Aron B, Sijbers J, Fieremans E (2016) Denoising of diffusion MRI using random matrix theory. Neuroimage 142:394-406.

Wong E (2012) Optimized phase schedules for minimizing peak RF power in simultaneous multi-slice RF excitation pulses. ISMRM 20th Annual Meeting. Melbourne, Australia.

Wright PJ, Mougin OE, Totman JJ, Peters AM, Brookes MJ, Coxon R, Morris PE, Clemence M, Francis ST, Bowtell RW, Gowland PA (2008) Water proton T1 measurements in brain tissue at 7, 3, and 1.5 T using IR-EPI, IR-TSE, and MPRAGE: results and optimization. MAGMA 21:121-130.
21

**SUPPLEMENTARY MATERIALS**

**Low-rank Reconstructions**

The low-rank matrix completion exploits the redundancy between the neighboring k-space data points. The linear dependency between the k-space points comes from the images with limited spatial support and/or slowly varying phase (Haldar, 2014). The procedure of low-rank reconstruction is illustrated in Figure S1. The reconstructed image is the solution of the equation in Figure S1. The equation is similar to the eq. (11) except that no estimated phase is involved and the solution includes the images of all the shots. The regularization parameter $\lambda$ is calculate as

$$\lambda = \|Y\|_2^2 / \|H(m_{ini})\|_* ,$$
(S1)

where $m_{ini}$ is the initial solution in eq. (S1) without the regularization term.

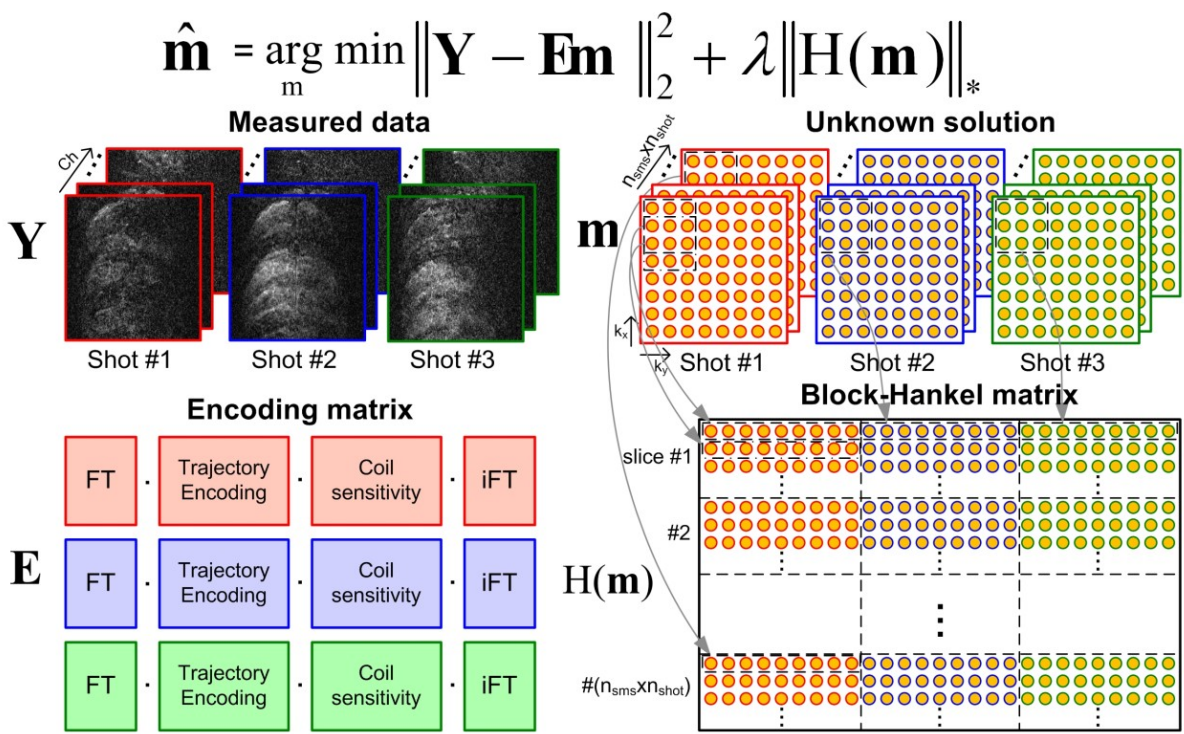

Figure S1: The illustration of low-rank reconstruction of MUSIUM data. **Y** denotes the measured data, **E** denotes the MUSIUM encoding matrix, **m** denotes the unknown image, $\lambda$ denotes the regularization paramter and **H**(·) denotes the Block-Hankel operation.

Figure S2a demonstrates the reconstruction results of eq. (11), which is SENSE-based. Figure S2b and S2c are both the results of low-rank reconstruction but with different regularization parameters. The image resolution is 0.86 mm isotropic and b value is 500. The regularization parameter λ in Figure S2b is calculated by eq. (S1) and the parameter in Figure S2c is 3λ. The SENSE-based images demonstrate less striping artifacts and better image contrast than those of low-rank images. Besides, Figure S2b and S2c demonstrate that the low-rank reconstruction is sensitive to the regularization parameter.

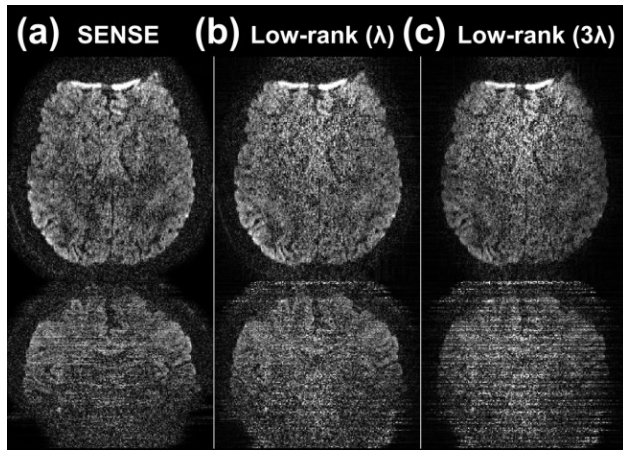

Figure S2: The reconstructed images of MUSIUM data. (a) The SENSE-based reconstruction. (b) The low-rank reconstruction with regularization parameter = λ. (c) The low-rank reconstruction with regularization parameter = 3λ.

**RFERENCES**